\documentclass[preprint,12pt]{elsarticle}

\newcommand{\bc}{\begin{center}}
\newcommand{\ec}{\end{center}}

\newcommand{\bd}{\begin{description}}
\newcommand{\ed}{\end{description}}

\newcommand{\bi}{\begin{itemize}}
\newcommand{\ei}{\end{itemize}}

\usepackage{graphicx}

\usepackage{amssymb}

\journal{Computer Physics Communications}

\begin{document}

\begin{frontmatter}



\title{\emph{Janus II}: a new generation application-driven computer for spin-system simulations}


\author[lab1,lab2,lab3]{M. Baity-Jesi}
\author[lab2,lab4]{R. A. Ba\~nos}
\author[lab4,lab2]{A. Cruz}
\author[lab1,lab2]{L. A. Fernandez}
\author[lab2]{J. M. Gil-Narvion}
\author[lab5,lab2]{A. Gordillo-Guerrero}
\author[lab2,lab11]{D. I\~niguez}
\author[lab3,lab2]{A. Maiorano}
\author[lab6]{F. Mantovani\fnref{filippo}}
\author[lab7]{E. Marinari}
\author[lab1,lab2]{V. Martin-Mayor}
\author[lab2,lab4]{J. Monforte-Garcia}
\author[lab1]{A. Mu\~noz Sudupe}
\author[lab8]{D. Navarro}
\author[lab7]{G. Parisi}
\author[lab2,lab11]{S. Perez-Gaviro}
\author[lab6]{M. Pivanti}
\author[lab7]{F. Ricci-Tersenghi}
\author[lab9,lab2]{J. J. Ruiz-Lorenzo}
\author[lab10]{S. F. Schifano}
\author[lab3,lab2]{B. Seoane}
\author[lab4,lab2]{A. Tarancon}
\author[lab6]{R. Tripiccione}
\author[lab3,lab2]{D. Yllanes}
\address[lab1]{Departamento de F\'isica Te\'orica I, Universidad Complutense, 28040 Madrid, Spain.}
\address[lab2]{Instituto de Biocomputaci\'on y Fisica de Sistema Complejos (BIFI),\\
 50009 Zaragoza, Spain.}
\address[lab3]{Dipartimento di Fisica, Universit\`a di Roma ``La Sapienza'', 00185 Roma, Italy.}
\address[lab4]{Departamento de F\'isica Te\'orica, Universidad de Zaragoza, 50009 Zaragoza, Spain.}
\address[lab5]{D. de Ingenier\'ia El\'ectrica, Electr\'onica y Autom\'atica,
U. de Extremadura,\\
10071 C\'aceres, Spain.}
\address[lab6]{Dipartimento di Fisica e Scienze della Terra, Universit\`a di Ferrara, and INFN,\\
44100 Ferrara, Italy.}
\fntext[filippo]{Now at Barcelona Supercomputing Center (BSC), 
08034 Barcelona, Spain.}
\address[lab7]{Dipartimento di Fisica, IPCF-CNR, UOS Roma Kerberos and INFN,\\
Universit\`a di Roma ``La Sapienza'', 00185 Roma, Italy.}
\address[lab8]{D. de Ingenier\'ia, Electr\'onica y Comunicaciones and I3A,
U. de Zaragoza,\\
50009 Zaragoza, Spain.}
\address[lab9]{Departamento de Fisica, Universidad de Extremadura, 06071 Badajoz, Spain.}
\address[lab10]{Dipartimento di Matematica e Informatica, Universit\`a di Ferrara, and INFN,\\
44100 Ferrara, Italy.}
\address[lab11]{Fundaci\'on ARAID, Diputaci\'on General de Arag\'on, Zaragoza, Spain.}

\begin{abstract}
This paper describes the architecture, the development and the implementation of
\emph{Janus II}, a new generation application-driven number cruncher optimized for Monte
Carlo simulations of spin systems (mainly spin glasses). This domain of
computational physics is a recognized grand challenge of high-performance
computing: the resources necessary to study in detail theoretical models that
can make contact with experimental data are by far beyond those available using
commodity computer systems. On the other hand, several specific features of the
associated algorithms suggest that unconventional computer architectures -- that
can be implemented with available electronics technologies -- may lead to order
of magnitude increases in performance, reducing to acceptable values on human
scales the time needed to carry out simulation campaigns that would take
centuries on commercially available machines. \emph{Janus II} is one such machine,
recently developed and commissioned, that builds upon and improves on the
successful JANUS machine, which has been used for physics since 2008 and is
still in operation today. This paper describes in detail the motivations behind
the project, the computational requirements, the architecture and the
implementation of this new machine and compares its expected performances with
those of currently available commercial systems.
\end{abstract}

\begin{keyword}
Spin glass \sep Monte Carlo \sep Application-driven computers \sep FPGA computing


\end{keyword}

\end{frontmatter}


\section{Overview}
\label{sec:overview}

Understanding glassy behavior is a major challenge in condensed matter physics
(see for instance Refs. \cite{ange,debene}). Glasses are materials that do not reach
thermal equilibrium on macroscopic time scales (e.g., years): bulk material
properties of a macroscopic sample, such as compliance modulus or specific
heat, change in time even if the sample is kept for days (years) at constant
experimental conditions. This {\em sluggish} dynamics is a major problem for
the theoretical and experimental investigation of glasses.
 
Spin glasses, usually regarded as prototypical glassy systems (or, more
generally, prototypical complex systems), have been extensively studied
theoretically; over the years this theoretical work has been widely supported by
numerical simulations, mostly using Monte Carlo techniques. The Monte Carlo
simulation of spin glass systems is a recognized grand challenge of computing,
as it requires inordinately large resources and at the same time it has
number-crunching requirements at large variance with
mainstream computer developments.

In a typical spin-glass model (see later for a detailed description), the
dynamical variables, the {\em spins}, are discrete and sit at the nodes of
discrete $D$-dimensional lattices. In order to make contact with experiments, we
may want to follow the evolution of a large enough 3$D$ lattice, say $100^3$
sites, for time periods of the order of 1 second. One Monte Carlo sweep (MCS)
---the update of all the spins in the lattice---  roughly corresponds to a time
scale of $10^{-12}$ seconds for a real sample, so we need some $10^{12}$ such
steps, that is $10^{18}$ spin updates. Also, in order to properly account for
disorder, we have to collect statistics on several (e.g., $O(100)$) copies of
the system, adding up to $10^{20}$ Monte Carlo spin updates. One easily reckons
that one needs a computer able to process on average one spin in 1
picosecond or less, in order to carry out this simulation program within
reasonable human timescales (say, less than one year).

The algorithms associated to the Monte Carlo simulation of spin glasses have
several properties that -- in principle -- open the way to very efficient
processing. First, as already noted, the degrees of freedom of several widely
studied spin glass models are discrete, and their values can be mapped on a
small number of bits (just one, for several popular models); discrete
bit-valued variables are operated upon with simple logic (as opposed to
arithmetic) operations, that can be performed by just a few logic gates.
Second, it is easy to identify a very large amount of parallelism in the
required computation, as one concurrently processes spins that do not have
direct interactions among one another.  Virtually all commercially available
computers are not able to exploit in full these properties; indeed, processors
are optimized for arithmetic (integer or -- even worse -- floating point)
operations, each such operation requiring a large number of logic gates. The
added burden of performing logic operations by hardware structures optimized to
perform arithmetic operations also severely limits the amount of parallel
computation that each processor is able to support. 

On the other hand,  these features, if consistently exploited, open the way to a
conceptually simple and efficient application-driven computing architecture,
carefully optimized for spin glass simulations, that promises to offer huge
performance advantages.  Application-oriented systems have been used in many
cases in computational physics, not only spin-system simulations but also in
Lattice QCD \cite{lqcd} and for the simulation of gravitationally coupled
\cite{grape} and biological \cite{anton} systems. Application-driven number
crunchers for spin systems have a long history:  the pioneering work by Pearson
and Richardson~\cite{PREVMACH_PRT} in the late 70s was followed by that of
Ogielsky and Condon~\cite{PREVMACH_OC} in the 80s; these pioneering attempts
were   followed by the SUE project \cite{sue} and more recently  by
JANUS \cite{JANUS_HW0, JANUS_HW1,JANUS_HW2} ---of which the work described in this paper is
the natural evolution. SUE and JANUS acknowledge that an
optimal architecture for spin simulators requires a dedicated processor
architecture and use Field Programmable Gate Arrays (FPGA) as the enabling
technology to implement that architecture.  FPGAs are integrated circuits that
can be configured at will after they have been assembled in an electronic
system.

In the last ten years, dedicated spin-glass crunchers, with their
order-of-magnitude better performance than available with traditional computers
have been instrumental to reach several key results in spin glass physics; the
most recent such machine JANUS, commissioned in 2008 and still in operation
today, has indeed made it possible to establish several new results
(see later for details).

In the same time frame, several innovative development in mainstream computer architecture -- including many-core processors and GPUs -- have made it possible to develop increasingly more parallel implementations of Monte Carlo algorithms for spin systems, significantly boosting performance, and largely closing the gap with respect to JANUS.
At the same time and in parallel with mainstream computer systems, progress in
electronic technology has also significantly boosted the level of parallelism
and performance that can be harvested using FPGAs.

This background has motivated the start of the development of \emph{Janus II}, described
in detail in this paper, that has the potential to provide order of magnitude
better performance than commercial computers in a time window of at least the
next five years, as well as superior energy efficiency.

\emph{Janus II} is an FPGA-based massively parallel spin-glass number cruncher, that
architecturally builds on JANUS and improves on it in several directions: i) it uses
latest generation FPGA technology, corresponding to  an order of magnitude
increase in performance per processor, ii) it includes an improved
communication interconnection among \emph{Janus II} nodes, that makes it efficient to
simulate large lattices using inter-node parallelism on top of intra-node
parallelism, iii) it enlarges by two orders of magnitude the size of the memory
available to the system and iv) it tightly couples the dedicated
number-cruncher nodes with traditional host computers, improving
data throughput and allowing a mixed-mode operation of the system in which
potentially complex control operations are handled efficiently by traditional
programs. All these improvements help boost the expected performance of \emph{Janus II}
as a spin glass number cruncher; moreover, points iii) and iv) above enlarge the
class of applications for which \emph{Janus II} is a potential efficient computer: while
the project is still mainly motivated as a spin glass simulator, we expect
interesting results is such diverse areas as graph theory, cryptography or
simulation of VLSI circuits.

This paper is structured in the following way: after this section, we
present a short introduction to spin glass models and to the Monte Carlo
techniques used to simulate them;  the paper continues with a description of the
\emph{Janus II} architecture, that closely matches the requirements outlined in the
preceding section. A  section on the programming and development environment
available for this machine follows, that also contains some performance
figures.  This is followed by a section that -- building on the expected
performance of the machine -- tries to identify several important questions in
spin glass physics accessible to \emph{Janus II} that were not within reach of JANUS. The
following section compares \emph{Janus II} performances with those available on currently
available computers and tries to forecast the extent of the window of
opportunity of our new machine. The paper ends with some concluding remarks.

\section{Spin Glass models}
\label{sec:SG}
Both JANUS and \emph{Janus II} have been designed from scratch to optimize their
performance for a specific application: the Monte Carlo simulation of spin
glasses. In this section we review the spin glass models that we want to study
with this machine.

Spin glasses are disordered magnetic alloys whose low-temperature phase is a
frozen disordered state, rather than the uniform patterns one finds in more
conventional magnetic systems~\cite{YOUNG,MYDOSH}. They are important because
they are widely regarded as the simplest possible model of a complex
system. In fact, as we will see below, spin glass models are extremely simple
to define.  In spite of this, finding the lowest energy configuration of a
three dimensional Ising spin glass is an
NP-hard problem~\cite{Barahona}. The main ingredients that make the problem so hard are
randomness in the interactions and {\em frustration}. By frustration we refer
to the impossibility of satisfying simultaneously all the demands that the
interactions pose on individual spins.

One of the most famous family of spin-glass models was proposed by Edwards and
Anderson~\cite{EA_MODEL} in the 70s. They consider a regular lattice and
define spins sitting at the lattice nodes. Spins are unit-length vectors of $n$
components: $\vec S_i=(S_{i,1},S_{i,2},\ldots,S_{i,n})$, 
$\vec S_i\cdot\vec S_i = 1$.  The interaction energy is
\begin{equation}
\mathcal{H} = - \sum_{i,j} J_{ij} \vec{S_i}\cdot \vec{S_j}\,,
\end{equation}
where the indices $i$ and $j$ run over the nodes of the lattice.
The \emph{coupling constants} $J_{ij}$ are chosen randomly (quenched disordered).  They are
statistically independent, and identically distributed. We shall be mostly
concerned with short-ranged interactions (i.e. $J_{ij}$ vanish, unless
$i$ and $j$ sites are lattice nearest neighbors). An instance of the coupling
constants $\{J_{ij}\}$ defines a {\em sample} of the physical system. The number $n$ of components of the
spins is also important. Some cases have special names:
Heisenberg ($n=3$), XY ($n=2$) and Ising ($n=1$).

The Ising spin glass model, $S_i=\pm 1$, with short-range interaction
(one of the prototypical material is  Fe$_{0.5}$Mn$_{0.5}$TiO$_3$) has
deserved special scientific attention for decades; its energy function reads:
\begin{equation}\label{eq:H-def}
\mathcal{H} = - \sum_{\langle i,j \rangle} J_{ij} S_i S_j \,,
\end{equation}
where $\langle i,j \rangle$ indicates that the sum runs only on
nearest neighbors in the lattice. Since the spin $S_i$ is a binary variable, it can be coded
on just one bit. Computational opportunities arise from this simplicity, and we
aim to explore some of them.

The goal of the game is to obtain assignments of the spin variables $\{S_i\}$ -- named {\em configurations} -- statistically distributed according to the
Boltzmann weight at temperature $T$:
\begin{equation}\label{eq:boltzmann}
P_{\mathrm B}(\{ S_i\}) = \frac{\mathrm{exp}[-\mathcal{H}(\{ S_i\})/T]}{Z}\,,\quad Z=\sum_{\{ S_i\}} \mathrm{exp}[-\mathcal{H}(\{ S_i\})/T]\,.
\end{equation}
One may try to achieve this goal by means of Markov Chain Monte Carlo
simulations (see, for instance, \cite{binder,HB,sokal97} for a detailed introduction). 
In principle, one just needs to implement some dynamics fulfilling
detailed balance and run it for a long enough time. However, in the limit of
vanishing temperature, the Boltzmann weight goes to zero unless the spin
configuration is a \emph{ground state} of the system (a lowest energy
configuration). Since finding ground states for a typical three dimensional spin-glass sample is a
NP-complete problem, something should go wrong with our simple-minded strategy.
The problem is in the length of the simulation: the autocorrelation times
for the Markov chain become inordinately large at low $T$; the simulation gets
trapped for a long time in some of the many local minima of the energy
(\ref{eq:H-def}). It is maybe worth mentioning that physical spin glasses (such
as Ag$_x$Mn$_{1-x}$, for instance) do suffer from the same problem: the system
does not reach thermal equilibrium even if it is allowed to evolve under
constant laboratory conditions for hours, or even days.

Finding equilibrium configurations for a single sample $\{J_{ij}\}$ is only half
of the problem. In order to obtain physically meaningful answers one needs to
average the thermal mean-values [i.e., the mean values corresponding to the
Boltzmann weight~(\ref{eq:boltzmann}) of a given sample], over a
fair number of samples [i.e. performing the average of the quenched
  disorder]. 
The meaning of \emph{fair} is very much dependent on the
physical questions that one asks and on the lattice size: it may range from
less than one hundred samples to maybe $10^5$ samples.

The dynamics that implement our Markov Chain Monte Carlo on a given sample at
temperature $T$ are pretty standard: Metropolis or Heat Bath. For instance,
the Metropolis procedure for the Ising spin glass starts from an initial
arbitrary configuration and generates new configurations  by picking one spin in
the lattice ($S_i$) and tentatively flipping it. One then computes the energy
difference $\Delta E$ associated to this tentative  change, $\Delta E = 2
\sum_{<j>} (J_{ij} S_i S_j)$ [where $j$ runs over all the nearest neighbors of
  the site $i$]. If $\Delta E \le 0$ the tentative flip is accepted
and the algorithm moves to another lattice site. If, on the other hand $\Delta E
> 0$, the tentative flip is accepted conditionally with a probability
proportional to $e^{- \Delta E \beta}$ ($\beta = 1/T$ is the inverse of the
system temperature).

One easily identifies a large degree of parallelism, as one
applies the procedure in parallel to any subset of spins that do not share a
coupling term in the energy function (so one correctly computes all $\Delta E$
terms): one usually partitions the lattice as a checkerboard and applies the
algorithm first to all black sites and then to all white ones, corresponding to
an available parallelism of degree $L^D/2$: in principle, we may schedule one
full Monte Carlo Sweep  (MCS, the application of the algorithm to all sites of the lattice) for any lattice size in just two computational steps, if enough
computational resources are available.

Simulations at constant temperature are not up to the task, if one wants to
produce a thermalized set of configurations at low temperature. One then resorts
to the parallel tempering (PT) algorithm~\cite{PT}. We consider $N_T$
temperatures $T_1<T_2<T_3<\ldots <T_{N_T}$. For each temperature, we consider a
statistically independent spin configuration $\{S_{i,a}\}$ with  $a=1,2,\ldots,
N_T$:
\begin{equation}
P_{\mathrm B}(\{ S_{i,a=1}\},\{ S_{i,a=2}\},\ldots, \{
S_{i,a=N_T}\}) = \prod_{a=1}^{N_T} \frac{\mathrm{exp}[-\mathcal{H}(\{ S_{i,a}\})/T_a]}{Z(T_a)}\,.
\end{equation}
Each of the $N_T$ systems is independently simulated at its own temperature by
means of one of the standard algorithms. However, every $n_{\mathrm{PT}}$ constant-$T$ sweeps,
one performs parallel tempering. The elementary parallel-tempering step is
the exchange attempt of the configurations at two consecutive temperatures
$T_a$ and $T_{a+1}$. The configuration exchange is accepted with Metropolis
probability:
\begin{equation}\label{eq:ProbPT}
\mathrm{Prob}_{\mathrm{PT}}=
\mathrm{min}\left[1,\frac{\mathrm{exp}\big(-\frac{\mathcal{H}(\{
      S_{i,a+1}\})}{T_a}-\frac{\mathcal{H}(\{
      S_{i,a}\})}{T_{a+1}}\big)}{\mathrm{exp}\big(-\frac{\mathcal{H}(\{
      S_{i,a}\})}{T_a}-\frac{\mathcal{H}(\{
      S_{i,a+1}\})}{T_{a+1}}\big)}\right] \,.
\end{equation}
One attempts to exchange configurations towards ascending temperatures
(so, in principle, the configuration at the lowest temperature could reach the
highest temperature in just one PT step): the rationale behind the parallel
tempering algorithm is simple. If a configuration trapped in a local minimum
is raised to a high enough temperature, it will be able to escape thanks to a
thermal fluctuation.

Parallel tempering has several tunable parameters. First, the
set of temperatures $\{T_a\}_{a=1}^{N_T}$ should be such that the acceptance
probability~(\ref{eq:ProbPT}) be reasonable (say, $\sim 10$\%). This
requires a relatively small temperature spacing. On the other hand, the
largest temperature $T_{N_T}$ should be high enough to ensure a quick
equilibration by means of the constant-$T$ algorithm. 
One has to reach a compromise among these conflicting goals,
as, the larger $N_T$, the larger are the needed computational
resources. The parameter that controls the parallel tempering frequency,
$n_{\mathrm{PT}}$, can be tuned as well. In our experience, the algorithmic
performance depends on $n_{\mathrm{PT}}$ only slightly. This is fortunate,
because the parallel tempering breaks for sure parallelism. One may
diminish its frequency (by increasing $n_{\mathrm{PT}}$), although some
tradeoff must also be found.

Let us finally mention that one may extend the Edwards-Anderson
model by including an external, site dependent magnetic field $h_i$:
\begin{equation}\label{eq:EA-extended}
\mathcal{H} = - \sum_{\langle i,j \rangle} J_{ij} s_i s_j - \sum_{i} h_i s_i\,.
\end{equation}
In this case, a sample is defined by the set of coupling constants
$\{J_{ij},h_i\}$. The addition of the local magnetic fields $h_i$ does not add
any real complication to the numerical simulation, and it has the advantages
of enlarging the set of problems that can be considered. Examples are the {\em
  random field Ising model (RFIM)} and the {\em diluted anti-ferromagnetic in a
  field model (DAFF)}~\cite{YOUNG}. A further extension consists on the
consideration of integer-valued spins $s_i=1,2,\ldots,Q$ (the so called
$Q$-states Potts model), which can be formulated in a similar way (see the
chapter by Binder in Ref. ~\cite{YOUNG}).

\section{\emph{Janus II} architecture}
\label{sec:archi}

The \emph{Janus II} architectural concept and its implementation follow directly from 
its predecessor (JANUS), built and commissioned in 2008 and still in operation 
today. The main guiding principle behind the old and new JANUS architectures 
is the attempt to leverage on state-of-the-art electronics technology in order to: 
i) exploit the huge parallelism available in the simulation of one spin glass (SG) system to 
speed up the Monte Carlo evolution of that system, ii) simulate in parallel a 
relatively large number of system samples, and iii) connect as tightly as possible 
the dedicated, massively parallel number crunching array with a traditional host 
computer system, so that complex and non-parallelizable computing functions 
(e.g., the proper handling of the parallel tempering temperature exchange) 
are done with as little  impact on global performance as possible.

The simulation of most SG models implies a mix of logic operations on bits  (as
opposed to arithmetic operations on long data words). Since virtually  all
commercially available computer architectures focus on arithmetic operations, 
they are conceptually a poor option for SG simulations.  An optimal choice would
be to hardwire all the logic gates that can be fabricated  on one silicon die in
order to perform exactly the set of required logic operations,  developing a
fully customized integrated circuit. This is possible in principle  (integrated
circuits designed to perform a specific function are called Application 
Specific Integrated Circuits, or ASICs), but the time and the costs
associated  to its development do not allow to pursue this option. 
We therefore choose the second best option, and adopt Field Programmable Gate
Arrays  (FPGAs) as the basic building block for \emph{Janus II}. FPGAs are integrated
circuits whose  logical gates can be connected at will, in order to perform a
specific set of logic  functions. FPGA configuration is a simple process that
can be done repeatedly, so  the same FPGA can be used for widely different logic
functions.  Currently available FPGAs have hundreds of thousands of so-called
\emph{logic cells},  each able to perform any logic operation of several bits; equally
important,  FPGAs come with several tens of Mbit embedded memory. 

\begin{figure} \centering
\includegraphics[width=0.85\textwidth]{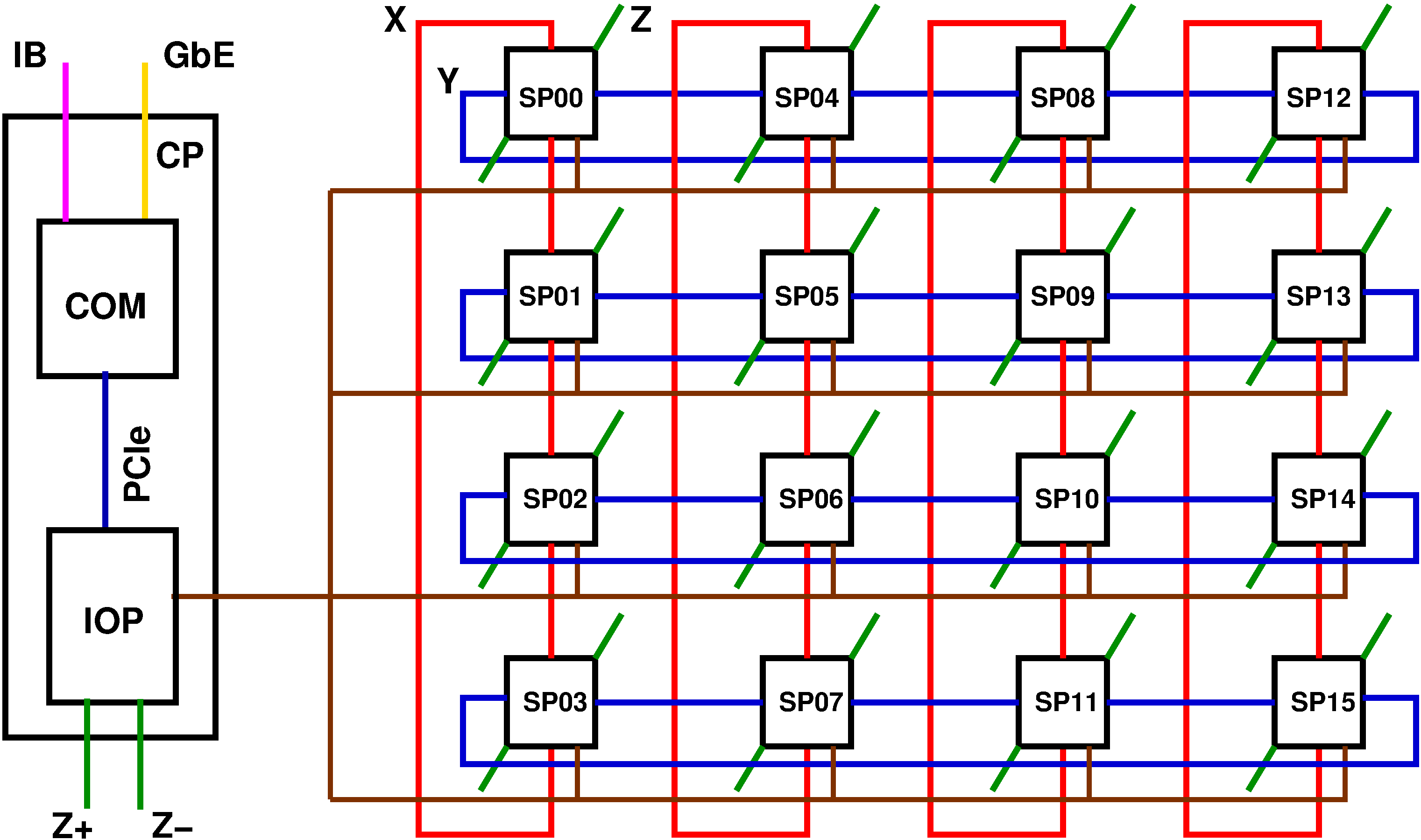}
\caption{\label{janus:arch} Architecture of the \emph{Janus II} Processing Board (PB). The array of 16 FPGA-based Simulation Processors (SPs, right) is connected by a 2$D$ ($x$ and $y$) toroidal network. All SPs have an additional independent connection to the IOP processor; the latter is part of the CP complex, that includes a commodity PC (adopting the COM form factor) and runs the Linux operating system; the CP has Gbit-Ethernet and Infiniband networking ports to the external world. Additional high speed connections are available for a tight coupling to other PBs in the $z$ direction.}
\end{figure}

The overall architecture of \emph{Janus II} is a parallel structure shown in figure~\ref{janus:arch}. 
The basic processing element of the system is the {\em Simulation Processor} (SP) whose 
computational structure is fully based on just one FPGA device. Each
SP includes one Xilinx Virtex-7 XC7VX485T FPGA and two banks of DDR-3 memory of
$8$ Gbyte each.
The choice of our FPGA has been done based mainly on cost and availability
issues  for this specific device. The selected FPGA has some $485000$ logic
cells and includes  $\sim 32$ Mbit embedded memory. As shown later in detail,
we expect  to embed within each SP more than 2000 spin-flip engines, each updating
one spin (all of the same color in a checkerboard structure) in one clock
cycle.  This corresponds to an average update rate of 1 spin every $2.5$ ps 
(with a conservative clock frequency of 200 MHz).

A set of 16 SPs are mounted onto a {\em Processing Board} (PB); the SPs of
each PB  are logically assembled at the nodes of a $4 \times 4$ array. Each SP
in the  array has direct point-to-point bi-directional links with its $4$
nearest neighbors; toroidal boundary conditions are applied. Each logical link
is engineered as 8 physical  links that we expect to operate at a bandwidth in
the range from $3$ to $5$ Gbit/second.

All SPs belonging to each PB are directly connected and controlled by a  Control
Processor (CP). The CP is a full fledged computer, running the Linux  operating
system. The CP plays  several roles in the \emph{Janus II} system: first, it is able to
configure the FPGAs of the  SPs, so they perform the desired logic operations; 
second it moves data from/to the SPs, so -- for instance -- initial data can be 
loaded to the SP and results of a simulation can go back to the CP. Finally, 
the CP controls the operations of all SPs, e.g. starting a simulation program, 
monitoring their status, collecting results, executing those parts of the global computation that cannot  be offloaded to the SPs and handling any errors. 

The CP uses a commercially available {\em Computer-on-Module} (COM) system,
based on an Intel Core i7 processor running at 2.2 GHz; it connects via the PCIe interface to a so-called
Input-Output-Processor (IOP) built inside yet another FPGA; the IOP actually 
manages all connections to all SPs, using a set of dedicated bi-directional high
speed links (one to each SP), running at  $\sim 3$ Gbit/s and a small number of
dedicated control and status lines. The IOP formats and appropriately routes data in transits from the CPU to the SPs, controls the configuration procedure of all SPs, controls their operation and monitors their status.
Since the IOP is itself a configurable unit, we are considering to use it -- on a longer time scale -- for additional computational/communication tasks; for instance the IOP might support a full crossbar switch among all SPs,  or handle directly the temperature exchange phase of a PT algorithm distributed over several SPs.

The CP is the main architectural improvement of \emph{Janus II}  with respect to its
predecessor: JANUS only had a Gigabit Ethernet link between  a set of SPs and
an external computer; the new arrangement increases the available  bandwidth
between the SP array and host to 4 Gbyte/second (a factor $\sim 40 \times$
larger than in the previous system)  and reduces  communication latency from
$\sim 15 \mu$s to $\sim 1 \mu$s. A much more  tightly coupled operation
of the SP array becomes possible, allowing to split  more finely a simulation
program on the control CPU and the SP array.

The combination of one CP and 16 SPs  is the basic functional block of a \emph{Janus II}
system. All these components are  assembled inside a box that also contains
power supplies and the forced-air  cooling system. This module operates as an
independent computing system and  can be networked with other \emph{Janus II} boxes and
with traditional computers via Ethernet and Infiniband interfaces.

A \emph{Janus II} installation can be made of any number $N$ of \emph{Janus II} boxes; the boxes
can be used as logically independent systems, running simulations of different
physical systems, or the whole system can operate as just one larger system; in
the latter case, the machine can be seen as a 3$D$ structure of $4 \times 4 \times
N$ SPs.  Bidirectional links are in fact available on each SP to build the
interconnection structure in the third dimension.

The project -- at the present stage -- has already assembled and tested a system with 16 \emph{Janus II} boxes, installed at BIFI in Zaragoza. The \emph{Janus II} team worked on the conceptual design of the system architecture while our industrial partner -- Link Engineering Srl, Bologna (Italy)~\cite{linkEng} -- have carried out the detailed engineering design and the actual construction of the prototype and of the presently available system.

\begin{figure}[t] \centering
\includegraphics[width=0.45\textwidth]{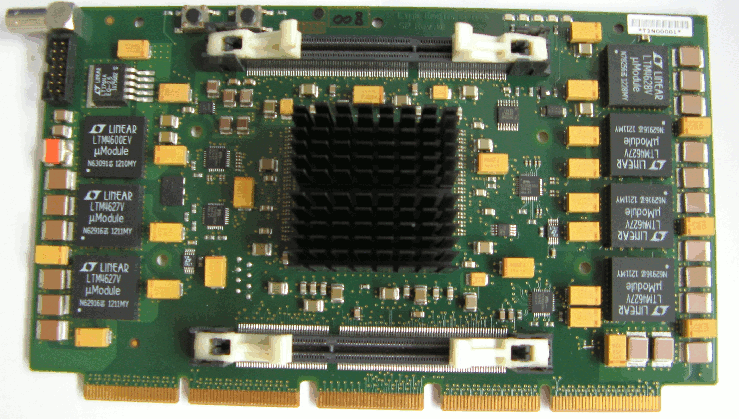}
\includegraphics[width=0.5\textwidth]{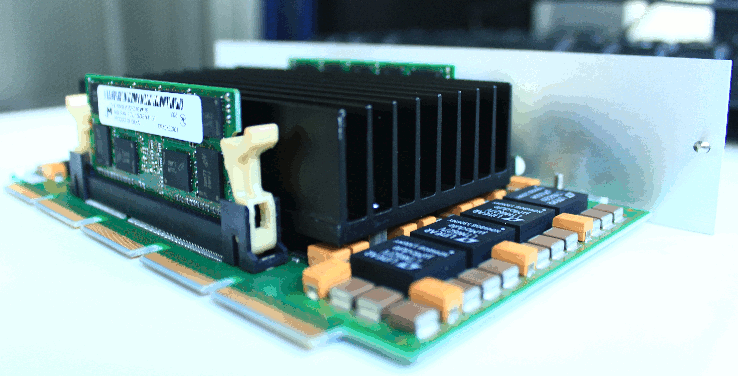}
\caption{\label{photo:sp} Pictures of a \emph{Janus II} SP module; the picture at left has a small heat radiator, providing a complete view of all components; the picture at right shows the large heat radiator needed to allow high frequency operation of the machine.}
\end{figure}

\begin{figure}[t] \centering
\includegraphics[width=0.80\textwidth]{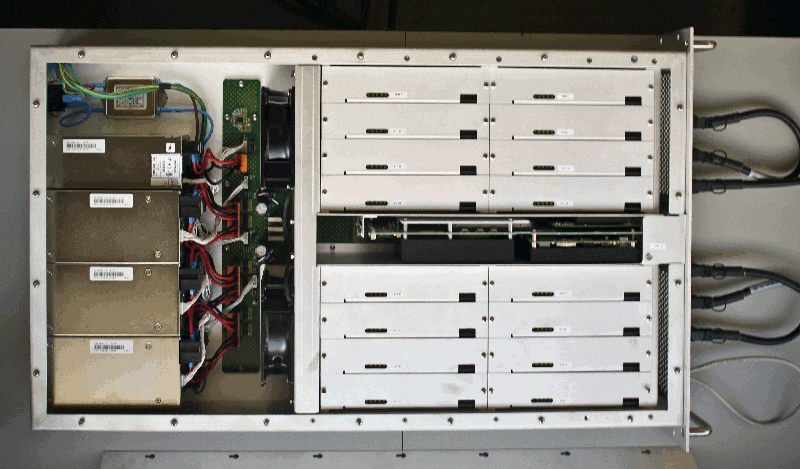}
\caption{\label{photo:box} Picture of a \emph{Janus II} box; there are 16 SP modules
(plugged vertically on the printed circuit, while the CP module is at the
center of the structure; at left one sees the cooling fans and the power
supplies.} 
\end{figure}

\begin{figure}[t] \centering
\includegraphics[width=0.7\textwidth]{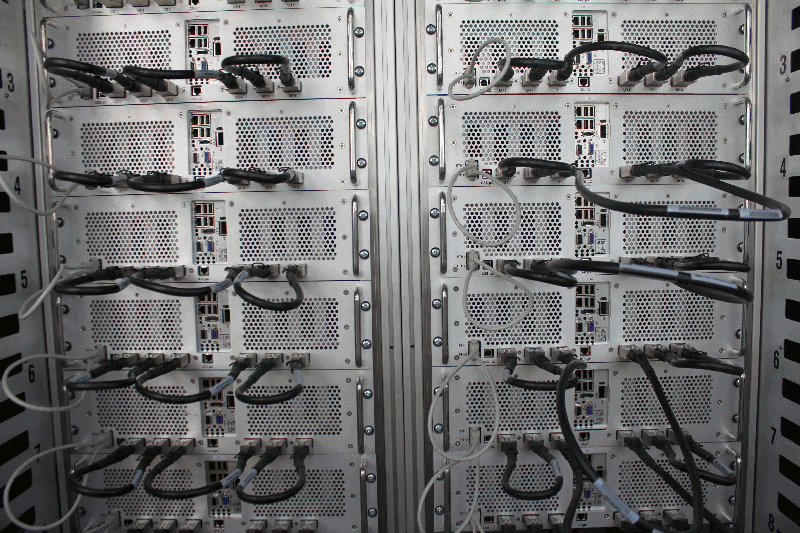}
\caption{\label{photo:rack} Close-up view of the \emph{Janus II} machine installed at BIFI (Zaragoza). The installation has 16 \emph{Janus II} boxes (12 are visible in the picture). The cables supporting the data-links in the $z$ direction are mounted in loop-back mode for test purposes.} 
\end{figure}

Fig.~\ref{photo:sp} shows an SP module while Fig.~\ref{photo:box} shows a
\emph{Janus II} box. Fig.~\ref{photo:rack} is a partial close-up view
of the fully assembled system.

\section{Structuring and programming a spin glass simulation on \emph{Janus II}}
\label{sec:progJanus2}
A \emph{Janus II} program is a combination of a standard C program, running on the CP and
a computational kernel, running on one or more appropriately configured SPs and
operating on data moved to the SP by the CP-resident program.  This programming
style is similar to the one usually adopted in processing systems that include
some form of co-processor or accelerator: a perhaps familiar example is GPU programming,
where the host processor sets up all required data-structures, initializes data
values and controls the outer loops of the program; the computationally heavy
kernels run on the GPU. The main difference is of course that, while GPUs
execute a program written in an appropriate programming language (e.g. CUDA or
openCL), SPs in \emph{Janus II} run the hardwired sequence of operations implied by the
configured FPGA.
Several development environments are available to assist in configuring FPGAs;
we use VHDL, a relatively low-level language that requires a  detailed
description of the structures that store data, the operations that are
performed on data and of instruction control: our experience shows however
that only this low level, largely handcrafted approach guarantees the high 
performance that we look for. 

From the perspective outlined in the previous paragraph, \emph{Janus II} might be seen as
a (possibly exotic) general purpose computer; however the main driving force
behind the project is of course that one expects  outstanding performance when
the SPs are configured for spin glass Monte Carlo simulations. Still, the fact
that \emph{Janus II} processing elements can be configured in arbitrary ways keeps the
door open for other uses of this machine.

The simplest operation mode for \emph{Janus II} will be the one already adopted for
JANUS: each SP performs a full Monte Carlo simulation of one SG system, while
different replicas of the system or physical systems at different temperatures are assigned to several SPs.

The update engine for one lattice site has a very simple structure. We consider again for definiteness the Ising spin glass in 3$D$; one maps the spins and coupling into bit-valued (\{0,1\}) variables:
\begin{equation}
S_k \rightarrow \sigma_k = (1 + S_k)/2 ~~~~~ 
J_{km} \rightarrow j_{km} = (1 + J_{km})/2\,.
\end{equation}
Once this is done, the evaluation of $\Delta E = 2 \sum_{<j>} (J_{ij} S_i S_j)$,
only implies 6 logic bit-wise {\tt xor} functions (replacing the products
$J_{ij} S_j$) followed by an arithmetic sum of just six bit-valued operands. The
result can be seen as the pointer to a small look-up table where the
corresponding pre-computed values of $e^{\beta \Delta E}$ are stored. At this
point, one arithmetically compares the value of the selected table entry with a
freshly generated random number: according to the outcome of the comparison the
previous value of the spin is left unchanged, or the flipped value is written to
memory. The required sequence of operation is similar for more complex spin
glass models or different Monte Carlo algorithms: different and (possibly) more
complex logic manipulations may be needed; in most cases the generation of
pseudo-random numbers remains the most complex operation. On JANUS we were able
to implement $\sim 1000$ such basic engines in each FPGA, using the
Parisi-Rapuano \cite{parisiRapuano} generator. With \emph{Janus II} we plan to
double this number and to increase the operating frequency by a factor $4$.
Under these conditions, the estimated power consumption of each SP -- based on
data made available by Xilinx -- is between $\sim 25$ and $30$ Watts.

One should notice that processing each spin implies reading $13$ bits and
writing one bit result (the new value of $S_i$) and reading a few 32-bit numbers
(3 for the Parisi-Rapuano generator) to compute the next element in the sequence
of random numbers. One quickly evaluates the overall memory traffic for 2000
spin-processing elements running at 200 MHz in excess of 4 Tbyte/second, orders
of magnitude beyond the bandwidth available with the large memory banks outside
the FPGA.
The needed bandwidth is on the other hand available using the large number of
memory blocks embedded {\em inside} our FPGAs; a rather complex memory
allocation scheme that matches our requirements and can be efficiently
implemented within the FPGA was devised for JANUS \cite{JANUS_HW1} and can be
carried over directly to \emph{Janus II}.  This requires however that all data items required by the program
fit inside the available on-chip memory. In our case the size of the FPGA
embedded memory is $\sim 32$ Mbit so we are able to handle 3$D$ lattices with $L
< 200$, taking into account that each lattice site needs 4 bits of data.
Alternatively, one can squeeze $30$ copies of a lattice of size $64^3$ inside
each SP, making it possible to run a large parallel tempering protocol on one or
two SPs. In this case, the CP would collect the energies of the lattices at all
temperatures $\{T_a\}$ after $n_\mathrm{PT}$ Monte Carlo steps, re-assign temperatures
according to Eq. (\ref{eq:ProbPT}) and start a new iteration.

If one wants to simulate larger lattices, all SPs can be used concurrently:
under the same assumptions as above, all $16$ SPs in one \emph{Janus II} box
are able to handle a 3$D$ lattice with $L \approx 500$ and even larger lattices
fit the complete array of 16 boxes; in this case, the lattice is partitioned
on all SPs in $1D$ or 2$D$ slices and data associated to abutting faces of the
sub-lattice are moved across SPs on the appropriate data links.

A combination of the strategies discussed produces extremely high computing performance
on \emph{Janus II}.  As discussed above,
we can partition the lattice on several SPs, slicing along one dimension.
The average time to process one spin on each processor is
\begin{equation}
T_\mathrm{spin} = \frac{1}{n_\mathrm{p} f} \,,
\end{equation}
where $n_\mathrm{p}$ ($n_\mathrm{p} \sim 2000$) is the number of update cores
available on each SP and $f$ is the SP operating frequency, expected in the
range from  125  to  250 MHz.

If we partition our lattice on $P$ processors (e.g., $P = 16$) the aggregated
mean spin update time is 
\begin{equation}
T_\mathrm{{global}} = \frac{1}{n_\mathrm{p} f P}\,,
\label{eq:tglobal} 
\end{equation}
corresponding to a $T_\mathrm{global}$  from  0.125 to  0.250 ps, in our frequency range.

In order to sustain these processing rates, the node-to-node communication
harness must provide a matching communication bandwidth: during the time in
which one SP updates all spins of its sublattice, we must move data associated to
the spin configuration of one face of the lattice from one SP to its neighbor.
Each SP sweeps all spins of its sublattice in a time
\begin{equation}
T_\mathrm{lat} = \frac{1}{n_\mathrm{p} f} ~ \frac{L^3}{P} \,.
\label{eq:tlat}
\end{equation}
The communication harness must move data belonging to one 2$D$ face of the
lattice in the same amount of time (this is just one bit per site on the
surface); assuming the network has $n_\mathrm{l}$ lanes each with a communication
bandwidth of $f_\mathrm{c}$ bit/second, we have
\begin{equation}
T_\mathrm{dat} = \frac{L^2}{n_\mathrm{l} f_\mathrm{c}} ~ = ~  
\frac{L^2}{n_\mathrm{l} (f_\mathrm{c}/f)} \frac{1}{f}.
\end{equation}

Communication is not a bottleneck as long as 
\begin{equation}
\frac{1}{n_\mathrm{p} } ~ \frac{L^3}{P} \ge \frac{L^2}{n_\mathrm{l} ~(f_\mathrm{c}/f)}\,.
\label{eq:balance}
\end{equation}

\begin{figure}
\centering
\includegraphics[width=0.9\textwidth]{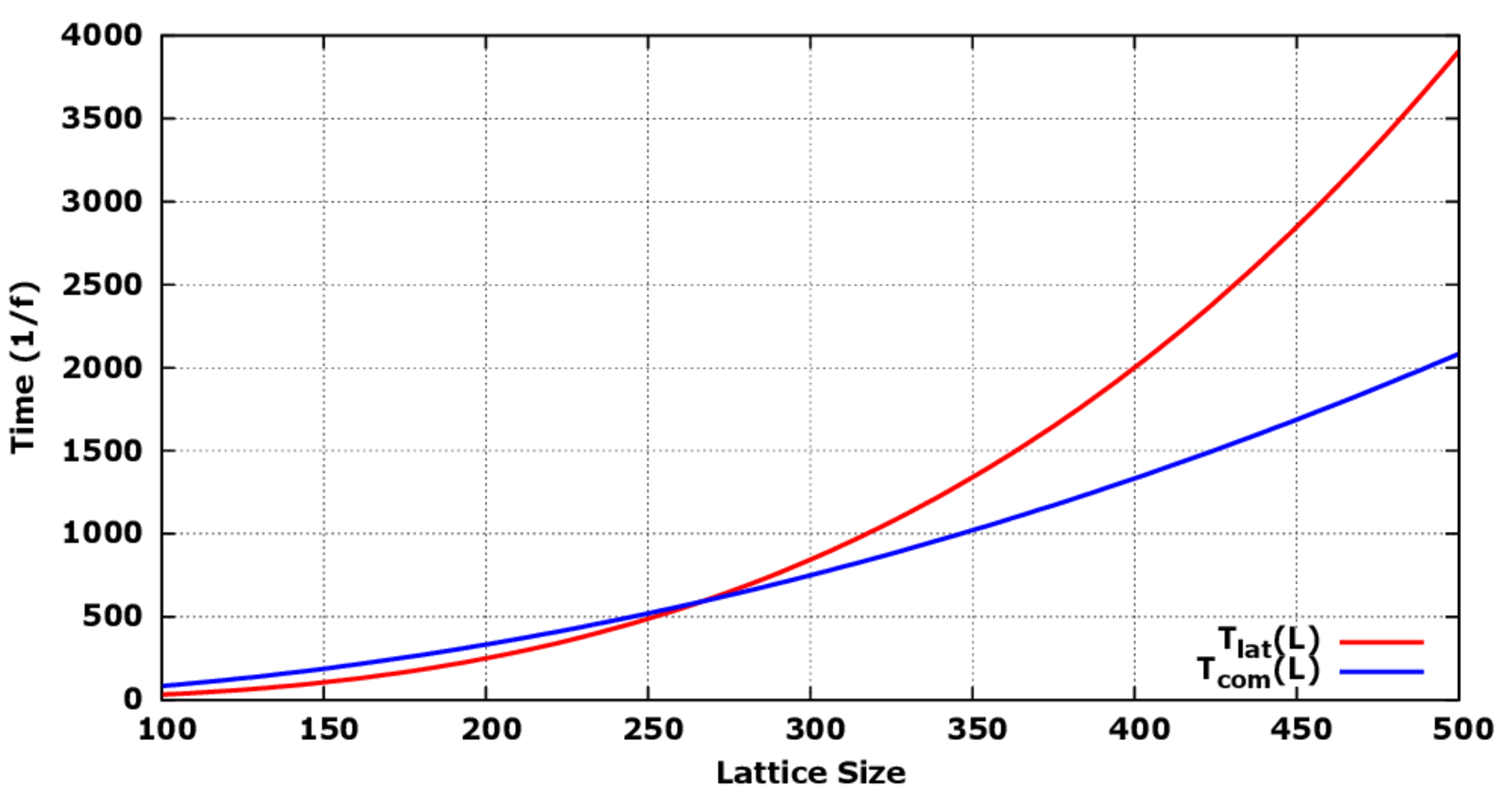}
\caption{Estimates of the computing time ($T_\mathrm{lat}(L)$, red) and the SP-to-SP
  communication time ($T_\mathrm{com}(L)$, blue) as a function of the lattice size
  $L$, assuming that the full lattice is split in $16$ strips, each assigned
  to one SP within a \emph{Janus II} box. One clearly sees that communication
  overheads are small for lattices of size $L \sim 150$ or larger and become
  fully negligible as soon as $L \ge 250$.}
\label{fig:balances}
\end{figure}

Figure \ref{fig:balances} shows the behavior of the two sides of Eq.
(\ref{eq:balance}) as a function of the lattice size $L$, with the already stated
values of the parameters and $f_\mathrm{c}/f = 15$ (we expect that $f_\mathrm{c}/f$ will be
somewhere in the $12$ to $20$ range): we see that the communication
infrastructure is powerful enough to handle lattices with $L \sim 250$ or
larger.

Let us consider a very large lattice for the current state-of-the-art (e.g., $L
= 500$); from either Eq. (\ref{eq:tglobal}) or Eq. (\ref{eq:tlat}) one finds
that the processing time for one sweep of the whole sublattice is of the order
of $T_\mathrm{proc}$ from  15 to 30 $\mu$s; in this simulation campaign, each \emph{Janus II}
box would run an independent replica of the system, so in one year of operation
one can hope to follow for several $10^{11}$ Monte Carlo steps of $\sim 10$
replicas of this very large system with  $3$ or 4 values of the temperature.

\section{\emph{Janus II} impact on spin-glass simulations}
\label{sec:challenges}
To a large extent, \emph{Janus II} is a follow up of JANUS, which has been a
major player in the field of spin glasses during the last five
years~\cite{JANUS1,JANUS2,JANUS3,JANUS4,JANUS5,JANUS6,JANUS7,JANUS8,epjst}. Hence,
it is natural to ask \emph{which are the important physics questions
accessible to \emph{Janus II} that were not within reach for JANUS?} 

In the previous sections we have estimated that the computing power available
from one SP in \emph{Janus II} is roughly $10\times$ larger than available
with JANUS. The (on board) available memory is also $10\times$ larger and,
last but not least, SP-to-SP communications make it possible to efficiently
simulate SG samples on just one or on a collection of SPs,
allowing flexible ways to trade the simulation speed of one sample with the
concurrent simulation of several samples.

Having these figures in mind, a rather blunt comparison with JANUS would be as follows. The total
number of spin updates in a simulation campaign is
\begin{equation}\label{eq:el-cuento-de-la-lechera}
N_\mathrm{spin-flips}=N_T\times N_\mathrm{spins}\times N_\mathrm{MCS}\times
N_\mathrm{samples}\,, 
\end{equation} 
where $N_T$ is the number of temperatures at which we simulate,
$N_\mathrm{spins}$ is the number of spins in the simulated lattice (i.e.,  in
$D$ spatial dimensions, for a lattice of size $L$, $N_\mathrm{spins}=L^D$),
$N_\mathrm{MCS}$ is the number of full-lattice updates performed for a single
sample and $N_\mathrm{samples}$ is the number of independent samples in the
simulation. As we said above, for a given wall-clock time, on \emph{Janus II} the l.h.s. 
of Eq.~(\ref{eq:el-cuento-de-la-lechera}) will be roughly ten times larger than on
JANUS. 

In fact, depending on the setup and the goals of the simulation campaign, with
\emph{Janus II} we can select which of the factors in~(\ref{eq:el-cuento-de-la-lechera})
we want to increase by $10\times$ or we can decide to spread the total gain on two or
more such factors. In addition, thanks to the improved communications, it is
possible to spread the simulation of a single sample over several FPGAs, thus
increasing further $N_\mathrm{spins}$ or $N_\mathrm{MCS}$ at the cost of
reducing $N_\mathrm{samples}$. It turns out that increasing by one order of
magnitude either $N_\mathrm{spins}$ or $N_\mathrm{MCS}$ or $N_\mathrm{samples}$
opens new opportunity windows. 

Roughly speaking, typical SG simulations come in two flavors:
\emph{non-equilibrium} and \emph{equilibrium\/}. Surprisingly enough, the two
turn out to be complementary~\cite{JANUS5}.

In non-equilibrium simulations one tries to analyze the
relaxation processes that take place in experimental spin glasses such as
CuMn. Below their glass temperature, such materials never reach thermal
equilibrium. Hence, one should perform simulations at a single temperature
(i.e.  $N_T=1$), with a dynamic rule such as Metropolis or heat-bath that
try to mimic the real spin dynamics. These simulations should be as long as
possible (i.e. $N_\mathrm{MCS}$ should be large), and the system
size (i.e. $N_\mathrm{spins}$) should be large enough to ensure that thermal
equilibrium is never approached. The only good news is that the number of
samples can be moderate, $N_\mathrm{samples}\sim 100$, because most of the
quantities that one computes are self-averaging (i.e., their sample-to-sample
fluctuations tends to zero as $1/N^a_\mathrm{spins}$, with $a\approx 1/2$).

On the other hand we have equilibrium simulations. Here, we need to approach
the equilibrium distribution, Eq.~(\ref{eq:boltzmann}). We are not tied to any
physical dynamics: any trick that one may invent is acceptable, provided that
it verifies the balance condition~\cite{sokal97}. In particular, we may employ
the parallel tempering algorithm explained in Sec.~\ref{sec:SG}, which
requires $N_T\sim 40$. As one may easily guess, the larger  the system size
the more valuable the physical information obtained from the simulation.
Unfortunately, the efficiency of parallel tempering is rather moderate:
JANUS established a world record by equilibrating lattices with $L=32$ in
three dimensions~\cite{JANUS4}. Another big issue is that the interesting
physical quantities are \emph{not} self-averaging at equilibrium:
sample-to-sample fluctuations are huge, which makes it desirable to simulate
a large number of samples.

At this point we are ready to appreciate the benefits of increasing by a
factor of 10 each of the individual factors in the r.h.s of
Eq.~(\ref{eq:el-cuento-de-la-lechera}).

\begin{itemize}
\item \emph{Increasing system sizes} will mostly benefit non-equilibrium
  simulations. Indeed, the coherence length $\xi(t)$, the typical size of the
  glassy domains, grows with the simulation time as $\xi(t)\sim t^{1/z(T)}$,
  with $z(T)\approx 6.86 \frac{T_\mathrm{c}}{T}$~\cite{JANUS1,JANUS2} (we
  measure the time $t$ in lattice sweeps; $T_\mathrm{c}=1.109(10)$ is the
  critical temperature~\cite{TC_HPV}). In experimental samples $\xi(t)$ is
  negligibly small as compared with the system size. Typical figures are
  $L=10^8$ and $\xi(t)\sim 100$ lattice spacings~\cite{Joh99,Bert04}. In fact,
  we know that in order to stay in the non-equilibrium regime one should have
  $L\geq 7 \xi(t)$~\cite{JANUS1}. In other words, for any $L$ there is a maximum
  safe simulation time $t^*$. This $t^*$ was amply surpassed in some of the
  simulations performed with JANUS. Indeed, in a month of continued operation
  one of the JANUS FPGAs simulated an $L=80$ lattice up to $t=10^{11}$ (this is
  the equivalent of one tenth of a second in physical time). However, in
  particular close to the critical temperature, $L=80$ is not large enough.
  Finite-size effects were felt at $t^*=10^9$. Fortunately, in the same month of
  continued operation \emph{Janus II} will be able to reach $t=10^{11}$ for
  lattice sizes $L=180$ (single FPGA), $L \simeq 500$ (16 FPGAs in a single
  board working in parallel) or $L \simeq 700$ (full machine). It is highly 
  unlikely that, for $t=10^{11}$ and $L \simeq 500$ finite-size effects will be
  relevant.

\item \emph{Increasing the number of samples}.  JANUS previous campaigns were
  remarkable for the sizes of the simulated samples, and the low temperatures
  reached. However, the number of simulated samples was typically in the range
  $1000-10000$. Some important physical effects, however, can be traced only
  through \emph{rare events.\/} Hence, an adequate investigation requires a
  significant boost in the number of samples (at least by a factor $\sim
  10$). There are at least two major problems where the sample-number issue is
  crucial. One is the so called temperature chaos problem
  \cite{Fernandez13}. The other is the survival (or lack of) of the spin glass
  phase in the presence of an external magnetic
  field~\cite{ATline_BM,ATline_YK,ATline_JK,JANUS7, JANUS8}. 

\item \emph{Increasing the simulation time}. Both equilibrium and
  non-equilibrium simulations may benefit by increasing
  $N_\mathrm{MCS}$. Non-equilibrium simulations for temperatures
  $T=0.6$ and below reached a quite modest coherence length $\xi(t)$ at
  $t=10^{11}$~\cite{JANUS1,JANUS2}. Thus, extending the duration of these
  $L=80$ simulations to $t=10^{12}$ will be informative while not endangering
  the non-equilibrium condition $L\geq 7 \xi(t)$. Another off-equilibrium
  example is the dynamical study of the possible transition
  in the physics of the spin glass in a field in $D=3$. With JANUS we were
  able to identify a dynamical transition, but our precision was not enough
  to decide definitively between several possible scenarios~\cite{JANUS8}.
  Extending the time window where we follow the evolution of the system could 
  be crucial to improve our understanding of this system.

 In equilibrium simulations
  one could either try to lower the reached temperature while keeping the
  system size fixed to $L=32$, or to increase the system size to $L=48$ while
  holding fixed the lowest temperature
  $T_\mathrm{min}=0.7026$~\cite{JANUS4}. By decreasing the lowest temperature,
  we could probe deeply in the spin-glass phase to study its many intriguing
  features (ultrametricity, statistics of overlap distributions, temperature
  chaos, etc.). On the other hand, increasing the system size at fixed
  temperature should allow us to assess finite-size effects, and to make
  rare-events less rare (the probability for a sample \emph{not} to display a
  rare event is expected to go as $\mathrm{exp}[-N_\mathrm{spins} \Omega]$
  with $\Omega$ small but positive~\cite{Fernandez13}).
\end{itemize}

Finally, let us mention another frontier to be explored, namely studying more
sophisticated spin glasses. Indeed, JANUS' limited memory implied that, in
practice, one was forced to consider only spin glass with Ising
spins. However, there are important problems~\cite{HSG_YS, AHSG_VS} that
cannot be treated within this framework. \emph{Janus II} should be able to simulate, at
the very least, XY spins [$n=2$ in Eq.~(\ref{eq:H-def})], maybe with some
discretization. In fact, a Migdal-Kadanoff renormalization study of the
discretization of the XY model by means of a clock model has recently
appeared~\cite{ilker13}. The discretization issue seems to be rather subtle
and worth of investigation on itself.

In short, the enhanced power of \emph{Janus II} will allow us to improve our understanding
of key topics in spin-glass physics that have already been investigated 
with JANUS (temperature chaos, ultrametricity, non-coarsening isothermal 
dynamics, presence of a phase transition in a field) but also to
delve into new problems (more sophisticated spin-glass models, non-isothermal dynamics, 
etc.).

\section{Performance comparison with commodity computers} 
When undertaking a major development project, like \emph{Janus II}, one should ensure
that the performance gain over commodity computers is large enough to justify
the effort and that this gap can be reasonably forecast to stay for a long
enough time window. In this section we compare the expected performance of
\emph{Janus II} with that available from several commodity systems, measured over the
last few years and try to derive reasonable forecasts for the near future.

We start reminding that the discussion of the previous sections shows that our
computational problem would optimally suit a {\em super-slim} processor that
handles bit-valued variables. Since commodity processors have wide data words
(and the current trend for recent processor is for wider and wider {\em vector}
words), efficient use of the computing resources mandates that spins and
couplings of different sites of the same lattice are grouped together on the
same (scalar or vector) data word and operated upon by bit-wise logic
operations; this approach -- that also naturally supports SIMD vectorization --
is known in the literature as multi-spin coding \cite{michael,bhanot}. One then
maps $V$ spins of a given sample on the same computer word and processes these
spins in parallel. In principle $V$ can be as large as the machine word size
$S$, but one independent random value is needed for each spin, so, as $V$
increases the incremental performance gain quickly fades away. As a further
optimization step, one can then process in parallel spins belonging to $W$ {\em
independent} samples (e.g., $W = S/V$) since just one random value can be used
to process $W$ spins belonging to independent samples, introducing a tolerable
amount of sample-to-sample correlation; in the following we will say that we
have a {\em sample} parallelism of degree $V$ and a {\em global} parallelism of
degree $W$. The optimal trade-off for most commercial architectures is that $V$
is significantly smaller than $S$, implying that a large number $W$ of samples
is simulated concurrently. This is useful from the point of view of accumulating
statistics over samples, but  -- we stress it once again -- in no way helps
solve the key problem of speeding up the Monte Carlo dynamics of {\em each}
sample; this is precisely where an application-driven architecture, for which 
$V$ is $O(10^3)$, produces its biggest dividends.

We will use performance metrics directly relevant for physics; we define the
{\em Sample spin update Time} (SUT) as the average time needed to update one
spin of one lattice sample. For each SP in \emph{Janus II} we have estimated in the
previous section  a SUT of 2 ps; for one full \emph{Janus II} box working on one lattice,
SUT goes down to 0.125 ps. We also define the {\em Global spin Update Time}
(GUT), appropriate when one simulation job handles several samples of the lattice
at the same time; GUT is simply defined as SUT/$W$. For \emph{Janus II}, GUT equals SUT
for each SP (and can be defined as SUT divided by the number of SPs working on
different samples).

When the JANUS project started, early 2006, state-of-the-art commodity
systems had dual-core CPUs; on those processors carefully optimized
codes had a SUT of $\sim 1000$ ps and GUT of  $\sim 400$ ps.
In the following  years, processors  have changed significantly with the
introduction of many-core CPUs and of general purpose GPUs; these
are better SG machines than traditional CPUs as one maps the available
parallelism on more cores (or on more threads, for GPUs).
%
%
\begin{table}[t]
\centering
\resizebox{\textwidth}{!}{
\begin{tabular}{||l|c|l|l|l|l|l|l|l|l|l||}
\hline
System                & Core 2 Duo & CBE (16 cores) & JANUS  & C1060 & NH (8 cores) & C2050 & SB (16 cores) & K20X & Xeon-Phi & \emph{Janus II} \\
\hline
Year                  & 2007	   & 2007	    & 2008   & 2009  & 2009	    & 2010  & 2012	    & 2012 & 2013     & 2013	 \\
Power (W)             & 150        & 220            & 35     & 200   & 220          & 300   & 300           & 300  & 300      & 25\\
SUT (ps/flip)         & 1000       & 150  	    & 16     & 720   & 200 	    & 430   & 60 	    & 230  & 52       & 2	 \\
Energy/flip (nJ/flip) & 150        &  33            & 0.56   & 144   & 244          & 129   & 18            & 69   & 15.6     & 0.05 \\
\hline
\end{tabular}
}
\caption{
Spin-update-time (SUT) of EA simulation codes on a $64^3$ lattice 
on several architectures. CBE is a system 
based on the IBM Cell processor; Tesla C1060, C2050 and K20X are NVIDIA 
GP-GPUs; NH (SB) are dual-socket systems based respectively on the 4-core 
Nehalem Xeon-5560 (8-core Sandybridge Xeon-E5-2680) processors, and 
Xeon-Phi is the recent launched MIC architecture of Intel. The table also shows rough estimates of the energy needed to perform all the computing steps associated to one spin flip.}
\label{comparison}
\end{table}
%
%
Over the years, we have compared~\cite{PPAM09,PARA10} JANUS with
several multi-core systems. In Table~\ref{comparison} we report the 
best SUT measured on several processors for a simulation of a lattice 
of $64^3$ sites.
We clearly see that over the years the large performance gap of 
Janus over commodity processor (e.g. Core 2 Duo) has been 
significantly reduced; an interesting first example was the 
extremely efficient IBM-Cell CPU, for which we have measured 
a SUT of $150$ ps. 
As of today, the best figure is offered by a 16-cores Sandy Bridge 
processor, for which SUT is $\approx 60$ ps.
Processor like the Xeon-Phi performs better on large lattices, for example we
have measured a SUT of 30 ps on a lattice of $128^3$ which improves the
performance of Sandy Bridge by a factor 2.  Equally significant is the energy
efficiency of the \emph{Janus II} system; data is shown again in Table \ref{comparison},
in which we display the approximate energy cost associated to the Monte Carlo update of one spin.


\begin{figure}[t]
\bc
\includegraphics[width=0.9\textwidth]{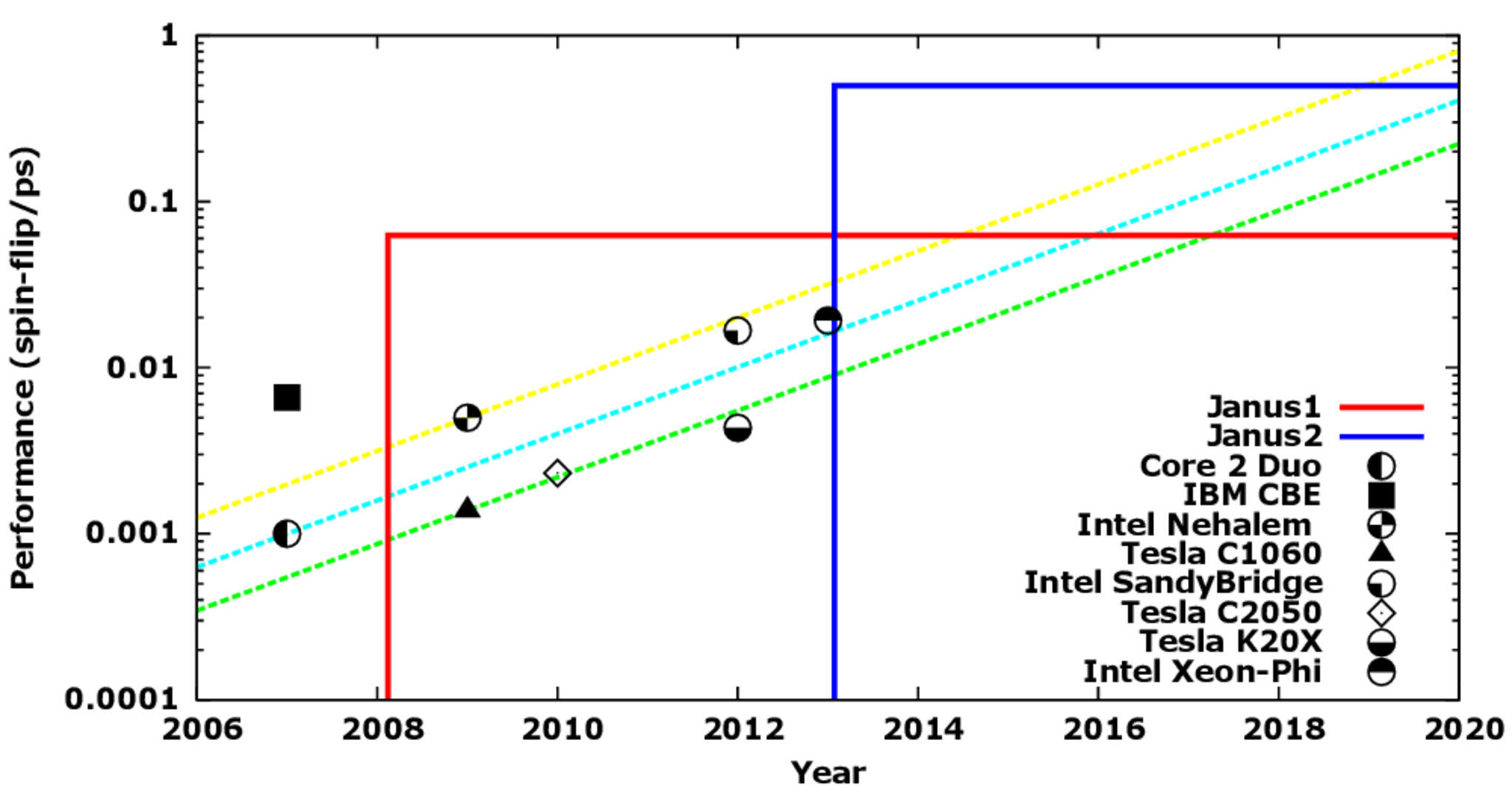}
\ec
\caption{Performance trends (measured in spin-flips/picosecond) for the simulation of the EA spin glass model with optimized programs 
for several commodity architectures and for JANUS and \emph{Janus II}. The lines
scale according to Moore's law. See the text for a complete discussion.}
\label{fig:moore}
\end{figure} 


All in all, a \emph{Janus II} box will be able to simulate in parallel a large spin glass
lattice more than 200 times faster than the best currently available commodity
option, and using $\sim 300$ times less energy. The next obvious question that
one has to face when developing a custom system is how long  it will keep its
performance edge over commercial systems.
Looking at Figure~\ref{fig:moore}, plotting data of Table~\ref{comparison}, we
see that performances of spin-glass applications on commodity systems have
increased over the time following a regular trend. Conversely
application-specific projects evolve in steps, as there is no performance
increase till a new generation is developed.
The plot clearly shows three lines of evolution of commodity systems: they all
scale according to Moore's law, with different pre-factors corresponding to
different broadly-defined families of architecture.
 
Looking at SUT figures for the Intel Nehalem and Sandy Bridge
micro-architectures with respect to those of the Core 2 Duo processor we
clearly see an abrupt jump in the scaling behavior associated to Moore's law;
we interpret this fact as the consequence of a performance gap that happened
when multi-core processors were introduced, followed by a regular Moore's
behavior (compare the two Moore's lines in the picture).
Looking at the performance plots  of the JANUS-class machines, we see that
JANUS will remain competitive until the end of 2014, and \emph{Janus II} comes into operation at the end-of-life of its predecessor; from this analysis we can reasonably look into our crystal-ball and expect that \emph{Janus II} should remain competitive through the year 2017.
Our analysis also shows the outstanding performance 
of the IBM-Cell processor, whose production has however been 
discontinued and the poor performance of GPU-based accelerator 
which suffer as they are more strongly optimized for floating-point arithmetics and lack cache-systems that are crucial for this class 
of applications.
Concerning the very recent Xeon-Phi processors, in spite of a 
very careful optimization, performances are not better than a dual Sandybridge 
system for small lattices; on the other hand, large on-chip caches in this processor keep its
performance constant on  larger lattices~\cite{hipc}.

\section{Conclusions}
\label{sec:conclusions}
In this paper we have described the architecture and implementation of the \emph{Janus II} application-driven machine, emphasizing its potential for performance in the simulation of spin glass systems. As described in detail in the previous section, the new machine will make it possible to carry out Monte Carlo simulation campaigns that would take centuries if performed on currently available computer systems.

The possibility to obtain such a large performance gap stems mainly from the
fact that the number crunching requirements associated to this class of
simulations are very different from those for which state-of-the-art computers
are optimized. At the same time, FPGAs offer an enabling technology that
allows to implement real machines with a reasonable engineering effort and at
costs affordable to a small scientific collaboration.

\emph{Janus II} builds and improves on the experience of its predecessor --
JANUS -- that has been running physics simulations for the last 6 years, and replaces the older machine at a point in time when the JANUS performance edge over commercial systems is significantly reduced.
JANUS and \emph{Janus II} have been designed with the main aim of
speeding up the Monte Carlo simulation of (a wide class of) spin glass
models. At the basic hardware level, both machines are not specialized for
these classes of simulations, so their use for other computational tasks is in
principle possible and efficient. In practice, attempts at using JANUS for
other applications have been hit by the serious bottleneck of the
small size of available memory. \emph{Janus II} addresses explicitly this
problem, since each SP node has 2 large banks of fast memory; we are now
starting to work on the assessment of the potential efficiency of our machine
for other applications, including such areas as cryptography, graph
optimization and simulation of VLSI circuits.

\section*{Acknowledgments}
We warmly acknowledge the excellent work done by the \emph{Janus II} team at
Link Engineering. In particular we thank  Pietro Lazzeri, Pamela Pedrini,
Roberto Preatoni, Luigi Trombetta and Alessandro Zambardi for their professional
and enthusiastic work.
The \emph{Janus II} project was supported by the European Regional Development Fund
(ERDF/2007-2013, FEDER project UNZA08-4E-020); by the European Research Council
under the European Union's Seventh Framework Programme (FP7/2007-2013, ERC grant
agreement no.  247328) ; by the MICINN (Spain) (contracts FIS2012-35719-C02,
FIS2010-16587); by Junta de Extremadura (contract GR101583); by the Italian
Ministry of Education and Research (PRIN Grant 2010HXAW77\_007).











\end{document}